\documentclass[jfm,aps,preprint,showpacs,preprintnumbers,amsmath,amssymb,secnumroman,eqsecnum]{revtex4}

\usepackage{graphicx}
\usepackage{dcolumn}
\usepackage{bm}

\newcommand{\beq}{\begin{equation}}
\newcommand{\eeq}{\end{equation}}
\newcommand{\beqa}{\begin{eqnarray}}
\newcommand{\eeqa}{\end{eqnarray}}
\newcommand{\vc}[1]{\mbox{\boldmath $#1$}}

\newcommand{\vol}[1]{{\bf #1}}

\begin{document}


\title{Swimming of a deformable slab in a viscous incompressible fluid with inertia}

\author{B. U. Felderhof}

 \email{ufelder@physik.rwth-aachen.de}
\affiliation{Institut f\"ur Theorie der Statistischen Physik\\ RWTH Aachen University\\
Templergraben 55\\52056 Aachen\\ Germany\\
}%

\date{\today}

\begin{abstract}
The swimming of a deformable planar slab in a viscous incompressible fluid is studied on the basis of the Navier-Stokes equations. A continuum of plane wave displacements, symmetric on both sides of the slab and characterized by a polarization angle, allows optimization of the swimming efficiency with respect to polarization. The mean swimming velocity and mean rate of dissipation are calculated to second order in the amplitude of the stroke. The optimum efficiency depends on the ratio of viscosity and mass density of the fluid. For high viscosity a stroke is found with significantly higher efficiency than Taylor's solution for a waving sheet. For low viscosity the efficiency is optimal for a nearly irrotational flow pattern.
\end{abstract}

\pacs{47.15.G-, 47.63.mf, 47.63.Gd, 45.50.Jf}
\maketitle
\section{\label{I}Introduction}

Historically the theory of swimming and flying has developed in two quite different directions \cite{1}. On the one hand there is the work on swimming and flying of a plate at high Reynolds number started by von K\'arm\'an and Burgers \cite{2} and continued by Lighthill \cite{3},\cite{4} and Wu \cite{5}-\cite{8}. The work is based on Euler's equations for irrotational flow of a fluid with vanishing viscosity. The discontinuity in velocity potential at the rear of the body is resolved by a vortex street. On the other hand there is the work on swimming at low Reynolds number based on Stokes equations for a viscous incompressible fluid. This work was started by Taylor \cite{9} with his model of a swimming sheet, and continued for bodies of different shapes by Lighthill \cite{10}, Purcell \cite{11}, and Shapere and Wilczek \cite{12},\cite{13} among others.

Taylor's work on a swimming sheet was extended by Reynolds \cite{14}, who considered the effect of fluid inertia on the swimming velocity. His work was corrected by Tuck \cite{15}, who calculated the mean rate of dissipation as well. For transverse plane wave displacements of given wavenumber and frequency the fluid flow to first order in the amplitude is irrotational in the limit of small viscosity, but the relation to the work of Lighthill and Wu is not clear. In the latter the swimming speed is treated as a parameter, and the thrust is calculated instead. The concept of thrust plays no role in the work of Reynolds and Tuck. The swimming of a sheet in the limit of small viscosity has been studied by Childress \cite {15A} by the technique of matched asymptotic expansions.

In recent work on swimming and flying of an assembly of rigid spheres \cite{16} we have developed a mechanical model for which the swimming velocity and mean rate of dissipation can be calculated in the whole range of fluid viscosity and mass density. In the model the mean thrust, equal and opposite to the mean drag, vanishes identically in periodic swimming. Further investigation into different models in the full range of viscosity and inertia would be helpful. In the following we study small amplitude swimming of a slab, which continues earlier work \cite{17} on the resistive limit. We begin by rederiving Tuck's expressions for a sheet using the method of Felderhof and Jones \cite{18}. Subsequently the same method is used to study the swimming of a slab.

The main purpose of the present study is to investigate the effect of fluid inertia on the efficiency of swimming for a simple model. Tuck's extension of Taylor's work on the swimming of a sheet showed that in that case fluid inertia has a negative effect in the sense that the swimming velocity for a stroke of given power is always less than for purely resistive Stokes flow. The same is true for swimming by longitudinal distortions in the plane of the sheet, also investigated by Tuck \cite{15}. On the other hand, for a slab one can consider swimming by potential flow, where viscosity plays no role and the swimming is achieved purely by inertia and the effect of added mass. As we show below, such potential swimming is just as efficient as that of Taylor's sheet or the squirming sheet in the Stokes limit.

After a discussion of Tuck's work on transverse waves on a sheet we study first the swimming of a slab by irrotational flow. The flow pattern has a different symmetry from that considered by Lighthill and Wu, the first order flow potential and pressure being continuous across the slab. The stroke consists of a circular motion of points on the surface of the slab, as in gravity waves, and of opposite circular polarization on both sides. There is no viscous boundary layer. It turns out that the swimming efficiency, defined as the ratio of swimming speed and mean rate of dissipation, divided by viscosity, is independent of both viscosity and fluid mass density. It has the same value as for Taylor's sheet.

Subsequently we study a squirming slab with displacements in the plane of the slab, so that the displacement wave is longitudinal. It turns out that the swimming efficiency for this stroke is always larger than that of Tuck's solution for the transverse wave. In both cases the efficiency tends to Taylor's value in the limit of large viscosity and to zero in the limit of small viscosity. The latter behavior implies that these linearly polarized waves give a poor description of swimming and flying at small dimensionless viscosity $\zeta$.

The wingtip path for birds is close to elliptic, or sometimes a figure eight \cite{20}. To obtain a better description of swimming and flying at small viscosity it is therefore natural to consider strokes with a forward motion in the form of an elliptically polarized wave. We consider in particular a slab with symmetric displacements on both sides, given by an elliptically polarized plane wave, characterized by a polarization angle $\alpha$. The displacement vector is given by Eqs. (2.12) and (6.1). For $\alpha=0$ the solution is identical to the squirming flow solution mentioned above. For $\alpha=\pi/4$ it corresponds to the potential flow solution, and for $\alpha=\pi/2$ it is identical to Tuck's transverse solution apart from a sign factor for the flow in the lower half-space.

The polarization angle $\alpha$ can be used to optimize the efficiency for fixed wavenumber and frequency. It turns out that in the resistive limit the efficiency is maximal for $\alpha=\pi/8$, with efficiency a factor $\sqrt{2}$ larger than for Taylor's solution, as we found earlier \cite{17}. For smaller values of the dimensionless viscosity $\zeta$ the optimum angle is in the range $\pi/8<\alpha<\pi/4$ and leads to a value of the optimum efficiency between one and $\sqrt{2}$ times that of Taylor's sheet. In the inertia-dominated limit, characterized by $\zeta\rightarrow 0$, the optimum corresponds to $\alpha=\pi/4$ and the potential flow solution. The optimum value of the efficiency in this limit is identical to that of Taylor's sheet. In Figs. 4 and 5 we present a sketch of the nature of the motion at the surface of the slab for this case. The calculation suggests that in other geometry as well, the potential flow solution without viscous boundary layer is of paramount importance in the limit of low viscosity.

\section{\label{2}Swimming slab}

We consider a planar slab immersed in a viscous incompressible fluid of shear viscosity $\eta$ and mass density $\rho$. The slab at rest is bounded by two planes at distance $2d$. We use Cartesian coordinates $x,y,z$ such that the upper $xz$ plane is at $y=d$ and the lower at $y=-d$. The rest shape of the slab is denoted as $S_0$. We shall consider a prescribed time-dependent shape $S(t)$ leading to swimming motion in the $x$ direction. The distortions are decomposed into $S_+(t)$ for the upper surface and $S_-(t)$ for the lower surface. The fluid is set in motion by the time-dependent distortions of the slab. The flow velocity
$\vc{v}(\vc{r},t)$ and the pressure $p(\vc{r},t)$ in the rest frame of the slab satisfy the
Navier-Stokes equations
\begin{equation}
\label{2.1}\rho\big[\frac{\partial\vc{v}}{\partial t}+(\vc{v}\cdot\nabla)\vc{v}\big]=\eta\nabla^2\vc{v}-\nabla p-\rho\frac{d\vc{U}}{dt},\qquad\nabla\cdot\vc{v}=0,
\end{equation}
where $\vc{U}(t)$ is the instantaneous swimming velocity.
We assume that the distortions are periodic in time with period $T=2\pi/\omega$. They give rise to a mean translational velocity $\overline{\vc{U}}=\overline{U}\vc{e}_x$ in the $x$ direction. The slab can be idealized to be infinite in the $x$ and $z$ directions. The flow velocity tends to $-\vc{U}(t)$ and the pressure tends to the ambient pressure $p_0$ as $y$ tends to $\pm$ infinity. We assume that the distortions of upper and lower surface have a symmetry such that they do not give rise to a rotational velocity.

The surface displacement
$\vc{\xi}(\vc{s},t)$ is defined as the vector distance
\begin{equation}
\label{2.2}\vc{\xi}=\vc{s}'-\vc{s}
\end{equation}
of a point $\vc{s}'$ on the displaced surface $S(t)$ from the
point $\vc{s}$ on the slab $S_0$. We decompose $\vc{s}$ into $\vc{s}_+=(x,d,z)$ for the upper surface and $\vc{s}_-=(x,-d,z)$ for the lower surface. The fluid
velocity $\vc{v}(\vc{r},t)$ is required to satisfy the no-slip boundary condition
\begin{equation}
\label{2.3}\vc{v}(\vc{s}+\vc{\xi}(\vc{s},t))=\frac{\partial\vc{\xi}(\vc{s},t)}{\partial t}.
\end{equation}
The displacement is decomposed into $\vc{\xi}_\pm(t)$ for the upper and lower surface. We consider displacements which do not depend on $z$ and take the form $\vc{\xi}_\pm=(\xi_{\pm x}(x,t),\xi_{\pm y}(x,t),0)$. As a consequence the flow velocity $\vc{v}$ and pressure $p$ do not depend on $z$, and the problem is effectively two-dimensional in the $x,y$ coordinates.

We construct an approximate perturbative solution to Eq. (2.1) with boundary condition (2.3) by formal expansion \cite{18} of the flow velocity and the pressure in powers of $\vc{\xi}$,
\begin{equation}
\label{2.4}\vc{v}=\vc{v}_1+\vc{v}_2+...,\qquad p=p_0+p_1+p_2+....
\end{equation}
Correspondingly the translational swimming velocity is expanded as
\begin{equation}
\label{2.5}\vc{U}=\vc{U}_1+\vc{U}_2+....
\end{equation}
The flow velocity at the surface is formally expanded as
\begin{equation}
\label{2.6}\vc{v}(\vc{s}+\vc{\xi}(\vc{s},t))=\vc{v}(\vc{s},t)+(\vc{\xi}\cdot\nabla)\vc{v}(\vc{r},t)\big|_{\vc{r}=\vc{s}}+....
\end{equation}
With the aid of this expansion the boundary condition may be applied at the undisplaced surface.

To first order the fluid equations of motion reduce to the linearized Navier-Stokes equations
\begin{equation}
\label{2.7}\rho\frac{\partial\vc{v}_1}{\partial t}=\eta\nabla^2\vc{v}_1-\nabla p_1-\rho\frac{d\vc{U}_1}{dt},\qquad\nabla\cdot\vc{v}_1=0.
\end{equation}
The first order boundary condition is
\begin{equation}
\label{2.8}\vc{v}_1(\vc{s})=\frac{\partial\vc{\xi}(\vc{s},t)}{\partial t},\qquad\vc{s}\in S_0.
\end{equation}

To second order the time-averaged flow $\overline{\vc{v}_2},\overline{p_2}$ satisfies the inhomogeneous Stokes equations  \cite{18}
\begin{equation}
\label{2.9}\eta\nabla^2\overline{\vc{v}}_2-\nabla\overline{p_2}=-\overline{\vc{F}_2},
\end{equation}
with force density
\begin{equation}
\label{2.10}\overline{\vc{F}_2}=-\rho\overline{(\vc{v}_1\cdot\nabla)\vc{v}_1},
\end{equation}
where the overline indicates averaging over a period $T=2\pi/\omega$. The force density may be written as the divergence of a Reynolds stress tensor. The Stokes equations Eq. (2.9) must be solved with the boundary condition
\begin{equation}
\label{2.11}\overline{\vc{v}}_2(\vc{s})=-\overline{(\vc{\xi}\cdot\nabla)\vc{v}_1}\;\big|_{\vc{s}},
\end{equation}
 as follows from Eq. (2.3). We shall consider situations where the value on the right is the same for $y=\pm d$, the mean flow velocity $\overline{\vc{v}_2}=\overline{v_2}\vc{e}_x$ depends only on $|y|$, and tends to a constant $-U_2\vc{e}_x$ as $y$ tends to $\pm\infty$. The prefactor $U_2$ can be identified as the second order mean swimming velocity of the slab. The force density $\overline{\vc{F}_2}$ is of the form $\overline{\vc{F}_2}=\overline{F_{2x}}(y)\vc{e}_x+\overline{F_{2y}}(y)\vc{e}_y$ and the mean pressure $\overline{p_2}$ depends only on $y$.

It will be convenient to use complex notation. Thus we write for the displacement vector of the upper and lower surface
\begin{equation}
\label{2.12}\vc{\xi}_\pm(x,t)=\vc{\xi}^c_\pm e^{ikx-i\omega t}
\end{equation}
with complex amplitude $\vc{\xi}^c_\pm$ and with the understanding that the real part of the expression is used to get the physical displacement. Correspondingly the first order velocity $\vc{v}_1(\vc{r},t)$ and pressure $p_1(\vc{r},t)$ take the form
\begin{equation}
\label{2.13}\vc{v}_1(\vc{r},t)=\vc{v}^c_\pm(y) e^{ikx-i\omega t},\qquad p_1(\vc{r},t)=p^c_\pm(y) e^{ikx-i\omega t}.
\end{equation}
The boundary value in Eq. (2.11) may be expressed as
\begin{equation}
\label{2.14}\overline{\vc{v}}_2(\vc{s})=-\frac{1}{2}\mathrm{Re}\;(\vc{\xi}^{*}_\pm\cdot\nabla)\vc{v}_1\big|_{y=\pm d},
\end{equation}
with values taken on either the upper or lower side of the slab.

The power required equals the rate of dissipation of energy in the fluid. This can be calculated from the work done per unit area against the stress $\vc{\sigma}=\eta(\nabla\vc{v}+(\nabla\vc{v})^T)-p\vc{I}$. The mean rate of dissipation per unit area is to second order \cite{18}
\begin{equation}
\label{2.15}\overline{D}_2=-\mathrm{Re}\;\vc{v}^{*}_1\cdot\vc{\sigma}_1\cdot\vc{e}_y\big|_{y=d}
\end{equation}
We have taken account of a factor 2 since there are equal contributions from the upper and lower half-space.

We define the dimensionless efficiency as
\begin{equation}
\label{2.16}E_2=4\eta\omega\frac{|U_2|}{\overline{D}_2}.
\end{equation}
We used the prefactor $4$ so that this equals unity for the case of a sheet in the resistive limit, as considered by Taylor \cite{9}.

\section{\label{III}Tuck's solution}

Tuck \cite{15} generalized Taylor's solution for an undulating thin sheet to the case of a fluid with inertia. Here we derive his results from Eqs. (2.9)-(2.11) and cast them in a more elegant form. We use the displacement vector Eq. (2.12) with
\begin{equation}
\label{3.1}\vc{\xi}^c_\pm=A\vc{e}_y
\end{equation}
with real amplitude $A$. In Fig. 1 we show a sketch of the slab with displacements given by Eq. (3.1). The first order flow velocity and pressure are given by Eq. (2.13) with
\begin{eqnarray}
\label{3.2}v^c_{x\pm}(y)&=&\mp B\omega\big(e^{\mp k(y\mp d)}-e^{\mp s(y\mp d)}\big) ,\qquad v^c_{y\pm}(y)=-iB\omega\big(e^{\mp k(y\mp d)}-\frac{k}{s}\;e^{\mp s(y\mp d)}\big),\nonumber\\
p^c_{\pm}(y)&=&\mp\frac{\omega^2\rho}{k}Be^{\mp k(y\mp d)},\qquad s=\sqrt{k^2-\frac{i\omega\rho}{\eta}},\qquad B=\frac{s}{s-k}A,
\end{eqnarray}
where the upper (lower) sign refers to the upper (lower) half-space.
In complex notation the force density $\overline{\vc{F}_2}$ in Eq. (2.10) can be expressed as
\begin{equation}
\label{3.3}\overline{\vc{F}_2}=-\frac{1}{2}\mathrm{Re}\;\rho(\vc{v}^*_1\cdot\nabla)\vc{v}_1.
\end{equation}
The $x$-component has the value
\begin{equation}
\label{3.4}\overline{F_{2x}}=\frac{1}{2}A^2\rho\omega^2\mathrm{Re}\;\frac{is}{|k-s|^2}\bigg((k+s) s^*e^{\mp(k+s) (y\mp d)}+k(k+s^*)e^{\mp(k+s^*)(y\mp d)}-k(s+s^*)e^{\mp(s+s^*)(y\mp d)}\bigg).
\end{equation}
The corresponding solution of Eq. (2.9) takes the form $\overline{\vc{v}_2}=\overline{v_{2x}}(y)\vc{e}_x$ with component
\begin{eqnarray}
\label{3.5}\overline{v_{2x}}(y)&=&-U_2\nonumber\\
&-&\frac{1}{2\eta}A^2\rho\omega^2\mathrm{Re}\;\frac{is}{|k-s|^2}\bigg(\frac{s^*}{k+s}\;e^{\mp(k+s) (y\mp d)}+\frac{k}{k+s^*}\;e^{\mp(k+s^*)(y\mp d)}-\frac{k}{s+s^*}\;e^{\mp(s+s^*)(y\mp d)}\bigg).\nonumber\\
\end{eqnarray}
From Eq. (2.14) we find the boundary condition
\begin{equation}
\label{3.6}\overline{v_{2x}}(\pm d)=\frac{1}{2}A^2\omega\;\mathrm{Re}\;s.
\end{equation}
Hence we find the swimming velocity
\begin{equation}
\label{3.7}U_2=-\frac{1}{2}A^2\omega k\;\mathrm{Re}f(\zeta)
\end{equation}
with dimensionless viscosity
\begin{equation}
\label{3.8}\zeta=\frac{\eta k^2}{\omega\rho}
\end{equation}
and the complex function
\begin{equation}
\label{3.9}f(z)=\frac{1}{2}+\sqrt{z(i-z)}.
\end{equation}
The expression in Eq. (3.7) can be shown to be identical to that derived by Tuck \cite{15},
\begin{equation}
\label{3.10}\mathrm{Re}f(\zeta)=\frac{1+F}{2F},\qquad F=\frac{1}{\sqrt{2}}\sqrt{1+\sqrt{1+\frac{1}{\zeta^2}}},\qquad\mathrm{for}\;\zeta>0.
\end{equation}
The variable $F$ tends to unity at large $\zeta$ and diverges as $1/\sqrt{2\zeta}$ at small $\zeta$. We note the identities
\begin{eqnarray}
\label{3.11}\sqrt{i+\zeta}&=&\sqrt{\zeta}\;F+\frac{i}{2\sqrt{\zeta}\;F},\nonumber\\
\sqrt{i-\zeta}&=&i\sqrt{\zeta}\;F+\frac{1}{2\sqrt{\zeta}\;F},\qquad\mathrm{for}\;\zeta>0.
\end{eqnarray}
Tuck calls $1/\zeta$ the Reynolds number. In Fig. 2 we plot $\mathrm{Re}f(\zeta)$ as a function of $\zeta$. For large $\zeta$ the function tends to unity in agreement with Taylor's result \cite{9}. For $\zeta\rightarrow 0$ the function tends to $1/2$.

We also consider the mean rate of dissipation. From Eq. (2.15) we find
\begin{equation}
\label{3.12}\overline{D}_2=A^2\eta\omega^2k\bigg(1+\mathrm{Im}\sqrt{\frac{i-\zeta}{\zeta}}\bigg).
\end{equation}
The viscous stress tensor does not contribute. It follows from Eq. (3.11) that the expression is identical to that derived by Tuck \cite{15},
 \begin{equation}
\label{3.13}\overline{D}_2=A^2\eta\omega^2k(1+F).
\end{equation}
It tends to Taylor's result $2A^2\eta\omega^2k$ as $\zeta$ tends to infinity, and it behaves as
\begin{equation}
\label{3.14}\overline{D}_2\approx A^2\sqrt{\frac{\eta\rho}{2}}\omega^{5/2}\qquad\mathrm {as}\;\;\eta\rightarrow 0,
\end{equation}
independent of the wavenumber. In Fig. 3 we plot the efficiency $E_2$, defined by Eq. (2.16), as a function of $\zeta$. The efficiency tends to unity for large $\zeta$ and vanishes in the limit $\zeta\rightarrow 0$.

It is of interest to consider also the kinetic energy of flow. We find in total per unit area
 \begin{eqnarray}
\label{3.15}\overline{\mathcal{K}_2}&=&A^2\frac{\rho\;\omega^2}{2k}\bigg(1+\sqrt{\zeta}\;\mathrm{Im}\sqrt{i+\zeta}\bigg)\nonumber\\
&=&A^2\frac{\rho\;\omega^2}{2k}\bigg(1+\frac{1}{2F}\bigg).
\end{eqnarray}
The expression shows that the mean virtual mass per unit area depends on the dimensionless viscosity. There is an interesting relation to the expression for $U_2$ in Eq. (3.7).

It follows from Eq. (3.2) that for small $\zeta$ the first order flow velocity consists of a potential flow and a thin boundary layer. In the limit $\zeta\rightarrow 0$ only the potential flow contributes to the kinetic energy. In this limit the swimming velocity tends to $U_2=-\frac{1}{4}A^2\omega k$, the mean rate of dissipation $\overline{D}_2$ vanishes, the mean kinetic energy tends to $\overline{\mathcal{K}_2}=A^2\rho\omega^2/(2k)$, and the efficiency vanishes, $E_2=0$. The mean kinetic energy is proportional to the mean thrust calculated in inviscid irrotational flow theory \cite{1},\cite{7}. In the present theory the concept of thrust does not appear. In the theory of inviscid flow end effects dominate and the swimming velocity is left undetermined. There are conceptual differences here that do not allow easy comparison.

\section{\label{IV}Potential flow}

In this section we compare the above results for the Taylor sheet with a second solution where the slab undulates in such a way that the flow is irrotational. In Taylor's solution the sheet undulates such that to first order the sheet is not extended. For the potential flow solution considered below the upper and lower surface of the slab are extended to first order. We use the displacement vector Eq. (2.12) with complex amplitude vector
\begin{equation}
\label{4.1}\vc{\xi}^c_\pm=\frac{A}{\sqrt{2}}\;(\vc{e}_x\pm i\vc{e}_y),
\end{equation}
corresponding to a circularly polarized wave. In Fig. 4 we show the displacements for a period. The first order flow velocity and pressure are given by Eq. (2.13) with
\begin{eqnarray}
\label{4.2}v^c_{x\pm}(y)&=&-i\omega\frac{A}{\sqrt{2}}\;e^{\mp k(y\mp d)},\qquad v^c_{y\pm}(y)=\pm\omega\frac{A}{\sqrt{2}}\;e^{\mp k(y\mp d)},\nonumber\\
p^c_{\pm}(y)&=&-i\frac{\omega^2\rho}{k}\frac{A}{\sqrt{2}}\;e^{\mp k(y\mp d)}.
\end{eqnarray}
Note that the first order pressure takes the same value on both sides of the slab, unlike Eq. (3.2).
From Eq. (2.14) we find the boundary condition
\begin{equation}
\label{4.3}\overline{v_{2x}}(\pm d)=-\frac{1}{2}A^2\omega\;k.
\end{equation}
Hence we find the swimming velocity
\begin{equation}
\label{4.4}U_2=\frac{1}{2}A^2\omega k,
\end{equation}
independent of viscosity. From Eq. (2.15) we find for the mean rate of dissipation
 \begin{equation}
\label{4.5}\overline{D}_2=2A^2\eta\omega^2k.
\end{equation}
The pressure does not contribute. For the efficiency defined in Eq. (2.16) we find $E_2=1$, independent of viscosity. For the kinetic energy of flow per unit area we find
\begin{equation}
\label{4.6}\overline{\mathcal{K}_2}=A^2\frac{\rho\omega^2}{4k},
\end{equation}
again independent of viscosity.

The expressions derived above show that the potential flow solution is quite simple, as it is for a distorting sphere \cite{19}. The swimming is more efficient than for Tuck's solution of Sec. III for any value of the dimensionless viscosity $\zeta$, as shown in Fig. 3. There is no viscous boundary layer in the first order flow. The mean swimming velocity is the same as for Taylor's sheet at the same mean rate of dissipation, but the mean kinetic energy of flow is one third of that of Taylor's sheet.

\section{\label{V}Squirming slab}

We consider also a squirming slab with equal plane wave displacements in the upper and lower plane of the form Eq. (2.12) with complex amplitude vector
\begin{equation}
\label{5.1}\vc{\xi}^c_\pm=A\vc{e}_x.
\end{equation}
In the limit of high viscosity the mean swimming velocity and the mean rate of dissipation are the same as for Taylor's sheet for the same amplitude of displacement $A$, but with swimming velocity in the opposite direction \cite{17},\cite{21}.

The first order flow velocity and pressure are given by Eq. (2.13) with
\begin{eqnarray}
\label{5.2}v^c_{x\pm}(y)&=&-iB\omega\big(e^{\mp k(y\mp d)}-\frac{s}{k}\;e^{\mp s(y\mp d)}\big) ,\qquad v^c_{y\pm}(y)=\pm B\omega\big(e^{\mp k(y\mp d)}-e^{\mp s(y\mp d)}\big),\nonumber\\
p^c_{\pm}(y)&=&-i\frac{\omega^2\rho}{k}Be^{\mp k(y\mp d)},\qquad s=\sqrt{k^2-\frac{i\omega\rho}{\eta}},\qquad B=\frac{k}{k-s}A.
\end{eqnarray}
The $x$-component of the mean force density in Eq. (3.3) has the value
\begin{equation}
\label{5.3}\overline{F_{2x}}=\frac{1}{2}A^2\rho\omega^2\mathrm{Re}\;\frac{ik}{|k-s|^2}\bigg((k+s) se^{\mp(k+s) (y\mp d)}+k(k+s^*)e^{\mp(k+s^*)(y\mp d)}-s(s+s^*)e^{\mp(s+s^*)(y\mp d)}\bigg).
\end{equation}
The corresponding solution of Eq. (2.9) takes the form $\overline{\vc{v}_2}=\overline{v_{2x}}(y)\vc{e}_x$ with component
\begin{eqnarray}
\label{5.4}\overline{v_{2x}}(y)&=&-U_2\nonumber\\
&-&\frac{1}{2\eta}A^2\rho\omega^2\mathrm{Re}\;\frac{is}{|k-s|^2}\bigg(\frac{s}{k+s}\;e^{\mp(k+s) (y\mp d)}+\frac{k}{k+s^*}\;e^{\mp(k+s^*)(y\mp d)}-\frac{s}{s+s^*}\;e^{\mp(s+s^*)(y\mp d)}\bigg).\nonumber\\
\end{eqnarray}
From Eq. (2.14) we find the boundary condition
\begin{equation}
\label{5.5}\overline{v_{2x}}(\pm d)=-\frac{1}{2}A^2\omega k.
\end{equation}
Hence we find the swimming velocity
\begin{eqnarray}
\label{5.6}U_2&=&\frac{1}{2}A^2\omega k\;\bigg(\frac{3}{2}-\sqrt{\zeta}\;\mathrm{Im}\sqrt{i+\zeta}\bigg)\nonumber\\
&=&\frac{1}{2}A^2\omega k\;\frac{3F-1}{2F}.
\end{eqnarray}
The second form was given by Tuck \cite{15}.

From Eq. (2.15) we find for the mean rate of dissipation
\begin{equation}
\label{5.7}\overline{D_2}=A^2\eta\omega^2 k\bigg(1+\frac{1}{\sqrt{\zeta}}\mathrm{Re}\sqrt{-i+\zeta}\bigg)=A^2\eta\omega^2 k(1+F).
\end{equation}
The mean kinetic energy of flow is
\begin{equation}
\label{5.8}\overline{\mathcal{K}_2}=A^2\frac{\rho\omega^2}{4kF}.
\end{equation}
In Fig. 3 we plot the efficiency $E_2$, defined by Eq. (2.16), as a function of $\zeta$. The efficiency tends to unity for large $\zeta$ and vanishes in the limit $\zeta\rightarrow 0$.

\section{\label{VI}Efficient swimming}

It is evident from Fig. 3 that of the three modes considered so far, the potential flow solution of Sec. IV leads to the most efficient swimming. In this section we show that for any value of the dimensionless viscosity $\zeta=\eta k^2/(\omega\rho)$ one can find a mode which is even more efficient. We consider intermediate modes depending on a polarization angle $\alpha$ which can be varied in such a way that the efficiency $E_2$ is optimized. The modes considered so far correspond to particular values of $\alpha$.

We consider plane wave displacements of the upper and lower plane of the form Eq. (2.12) with complex amplitude vector
\begin{equation}
\label{6.1}\vc{\xi}^c_\pm=A\cos\alpha\;\vc{e}_x\pm iA\sin\alpha\;\vc{e}_y,
\end{equation}
corresponding to an elliptically polarized plane wave.
For $\alpha=0$ this corresponds to the squirming displacement of Sec. V. For $\alpha=\pi/4$ it corresponds to the potential flow solution of Sec. IV. For $\alpha=\pi/2$ it corresponds to Tuck's solution of Sec. III with displacement and flow in the lower half-space differing by a minus sign. The sign change corresponds to a shift of the flow pattern by half a wavelength in the $x$ direction and does not affect the mean second order force density or flow velocity. The first order flow velocity and pressure are found by linear combination of the expressions in Eqs. (3.2) and (5.2).

The $x$-component of the mean force density in Eq. (3.3) takes the form
\begin{equation}
\label{6.2}\overline{F_{2x}}=\frac{1}{2}A^2\rho\omega^2\mathrm{Re}\big[\overline{f_{2x}}_1(1-\sin 2\alpha)+\overline{f_{2x}}_2\cos 2\alpha\big],
\end{equation}
with coefficients given by
\begin{eqnarray}
\label{6.3}\overline{f_{2x}}_1&=&\frac{i}{2|k-s|^2}\bigg(s|k+s|^2e^{\mp(k+s) (y\mp d)}+k|k+s|^2e^{\mp(k+s^*)(y\mp d)}-2ks(s+s^*)e^{\mp(s+s^*)(y\mp d)}\bigg),\nonumber\\
\overline{f_{2x}}_2&=&\frac{i}{2|k-s|^2}\bigg(s(k+s)(k-s^*)e^{\mp(k+s) (y\mp d)}+k(k-s)(k+s^*)e^{\mp(k+s^*)(y\mp d)}\bigg).
\end{eqnarray}
Evidently for $\alpha=\pi/4$ the expression in Eq. (6.2) vanishes, in accordance with Sec. IV. From Eq. (6.2) one evaluates the second order flow component $\overline{v_{2x}}(y)$ as in Eqs. (3.5) and (5.4). From Eq. (2.14) we find the boundary condition
\begin{equation}
\label{6.4}\overline{v_{2x}}(\pm d)=-\frac{1}{4}A^2\omega\;\mathrm{Re}\big[k-s+(k+s)\sin2\alpha+(k+s)\cos2\alpha\big].
\end{equation}
Hence we find the swimming velocity
\begin{equation}
\label{6.5}U_2=\frac{1}{2}A^2\omega k\;\bigg[\frac{1}{2}-\frac{1}{2F}+\bigg(\frac{1}{2}+\frac{1}{2F}\bigg)\sin2\alpha+\cos2\alpha\bigg].
\end{equation}

Similarly we find for the mean rate of dissipation
\begin{equation}
\label{6.6}\overline{D_2}=A^2\eta\omega^2k\;\big[1+F+\big(1-F\big)\sin2\alpha\big],
\end{equation}
and for the mean kinetic energy of flow
\begin{equation}
\label{6.7}\overline{\mathcal{K}_2}=A^2\frac{\omega^2\rho}{2k}\;\bigg[\sin^2\alpha+\frac{1}{2F}(1-\sin2\alpha)\bigg].
\end{equation}

The mean rate of dissipation $\overline{D_2}$ has a minimum at $\alpha=\pi/4$, corresponding to potential first order flow, independent of the value of $\zeta$. The ratio $|U_2|/\overline{D_2}$ yields the efficiency $E_2$ according to Eq. (2.16). For fixed value of $\zeta$ this can be maximized by variation of the polarization angle $\alpha$. In the limit of large $\zeta$ the maximum is at $\alpha=\pi/8$ and then the efficiency takes the value $\sqrt{2}$. For large $\zeta$ the mean kinetic energy of flow $\overline{\mathcal{K}_2}$ is minimal at $\alpha=\pi/8$. For decreasing $\zeta$ the maximum of $E_2$ shifts to larger values of $\alpha$, tending to $\pi/4$ as $\zeta\rightarrow 0$. At all values of $\zeta$ the maximum is larger than unity, and it tends to unity as $\alpha\rightarrow\pi/4$. For optimal swimming at small $\zeta$ the flow field is nearly irrotational. In Fig. 4 we plot the distortions of the slab for purely irrotational swimming. We have assumed a symmetry between the upper and lower half space, but the flow pattern in the lower half-space can be shifted horizontally without change in swimming speed or rate of dissipation. In particular, the distortions of the slab look like those shown in Fig. 5 for a shift by half a wavelength.

\section{\label{VII}Discussion}

The expressions derived in Sec. VI show that for symmetric plane wave displacements of a slab the swimming performance depends in a complex way on viscosity and mass density of the fluid and on the polarization angle characterizing the stroke. The dependence allows optimization of the swimming efficiency by variation of the polarization angle. In the resistive limit there is a stroke whose  efficiency is larger by a factor $\sqrt{2}$ than that of Taylor's sheet. In the inertial limit the optimal solution is found for nearly irrotational flow.

The calculation of mean swimming velocity and mean rate of dissipation has been performed to second order in the amplitude of displacements. Clearly it would be of interest to carry the calculation to higher order, but it may be expected that the present calculation provides a good picture of the qualitative trend.

The calculation is purely kinematic in the sense that the surface displacements are prescribed. A more elaborate calculation would take account of the elastic properties of the slab. The swimming as a consequence of an actuating force density could then be studied in a dynamic context. Also it would be of interest to extend the calculation to a compressible fluid.

\newpage

\section*{Figure captions}

\subsection*{Fig. 1}
Plot of the slab with plane wave distortion corresponding to Tuck's solution discussed in Sec. III. A cross-section in the $xy$ plane with a length of one wavelength is shown. The motion does not depend on the $z$ coordinate.

\subsection*{Fig. 2}
Plot of the function Re$f(\zeta)$ given by Eq. (3.10). The mean swimming velocity in Tuck's solution is proportional to this quantity, as shown in Eq. (3.7).

\subsection*{Fig. 3}
Plot of the efficiency $E_2$ as a function of dimensionless viscosity $\zeta=\eta k^2/(\omega\rho)$ for Tuck's solution (solid curve), for the squirming slab (long dashes), and for the irrotational flow solution (short dashes).

\subsection*{Fig. 4}
Plot of the slab with symmetric plane wave distortion corresponding to the potential flow solution discussed in Sec. IV. A cross-section in the $xy$ plane with a length of one wavelength is shown. The motion does not depend on the $z$ coordinate.

\subsection*{Fig. 5}
Plot of the slab with plane wave distortion corresponding to the potential flow solution discussed in Sec. IV with the wavepattern for the lower half-space shifted in the $x$ direction by half a wavelength in comparison with Fig. 4.

\newpage
\setlength{\unitlength}{1cm}
\begin{figure}
 \includegraphics{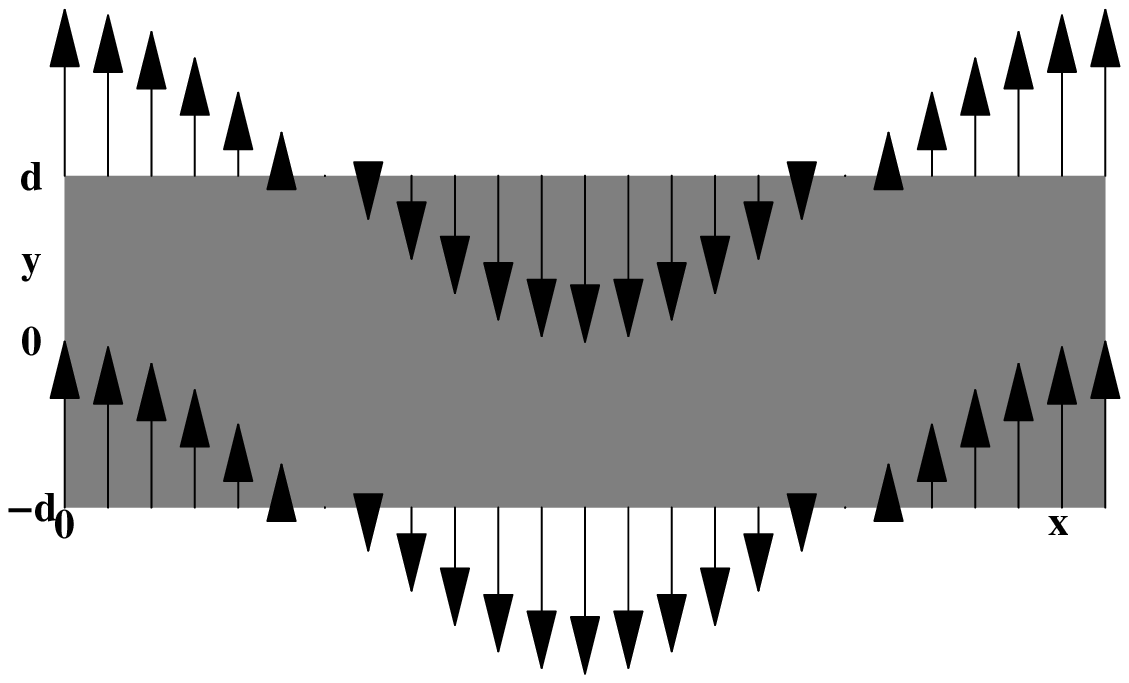}
   \put(-9.1,3.1){}
\put(-1.2,-.2){}
  \caption{}
\end{figure}
\newpage
\clearpage
\newpage
\setlength{\unitlength}{1cm}
\begin{figure}
 \includegraphics{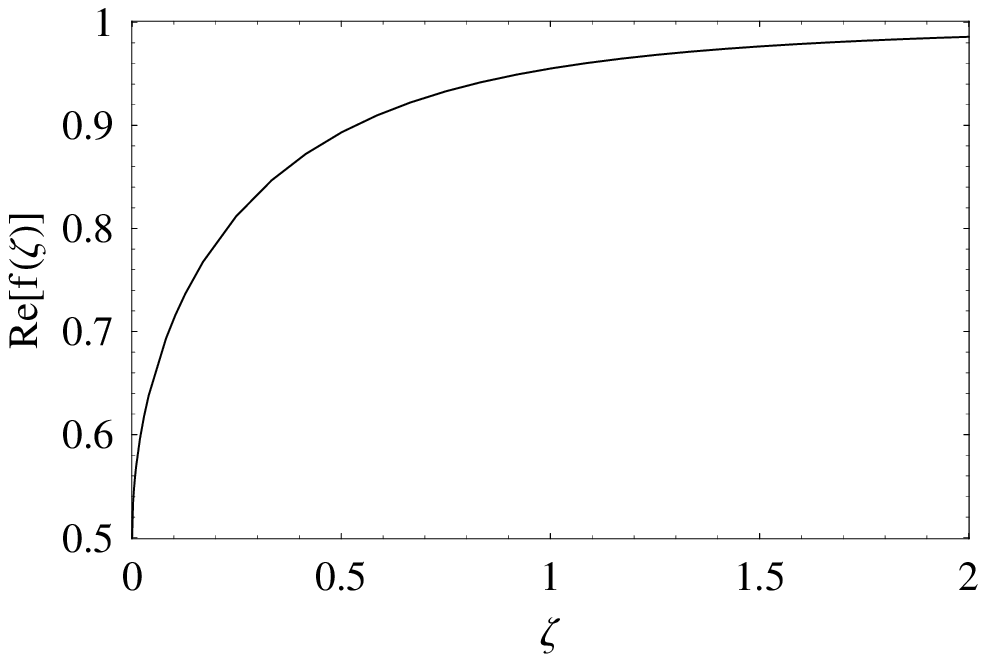}
   \put(-9.1,3.1){}
\put(-1.2,-.2){}
  \caption{}
\end{figure}
\newpage
\clearpage
\newpage
\setlength{\unitlength}{1cm}
\begin{figure}
 \includegraphics{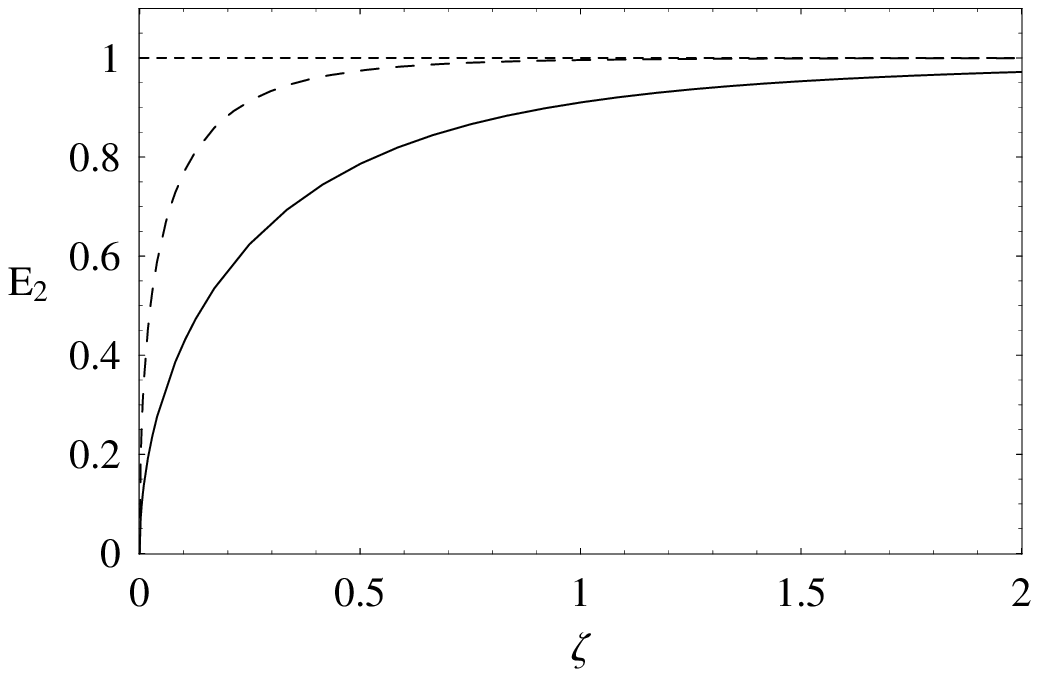}
   \put(-9.1,3.1){}
\put(-1.2,-.2){}
  \caption{}
\end{figure}
\newpage
\clearpage
\newpage
\setlength{\unitlength}{1cm}
\begin{figure}
 \includegraphics{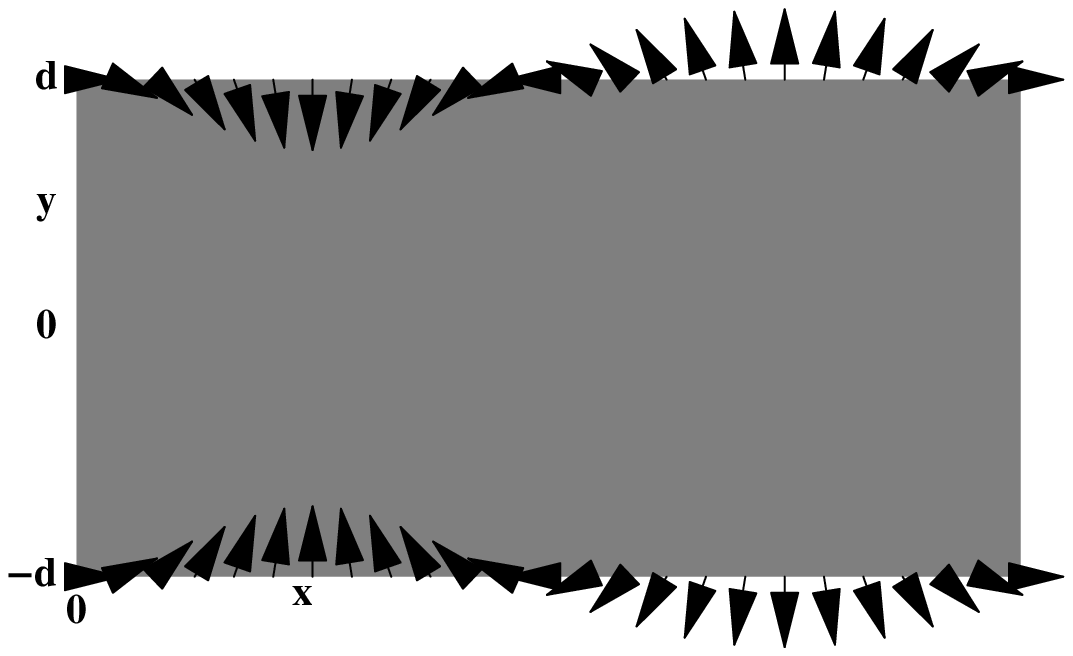}
   \put(-9.1,3.1){}
\put(-1.2,-.2){}
  \caption{}
\end{figure}
\newpage
\clearpage
\newpage
\setlength{\unitlength}{1cm}
\begin{figure}
 \includegraphics{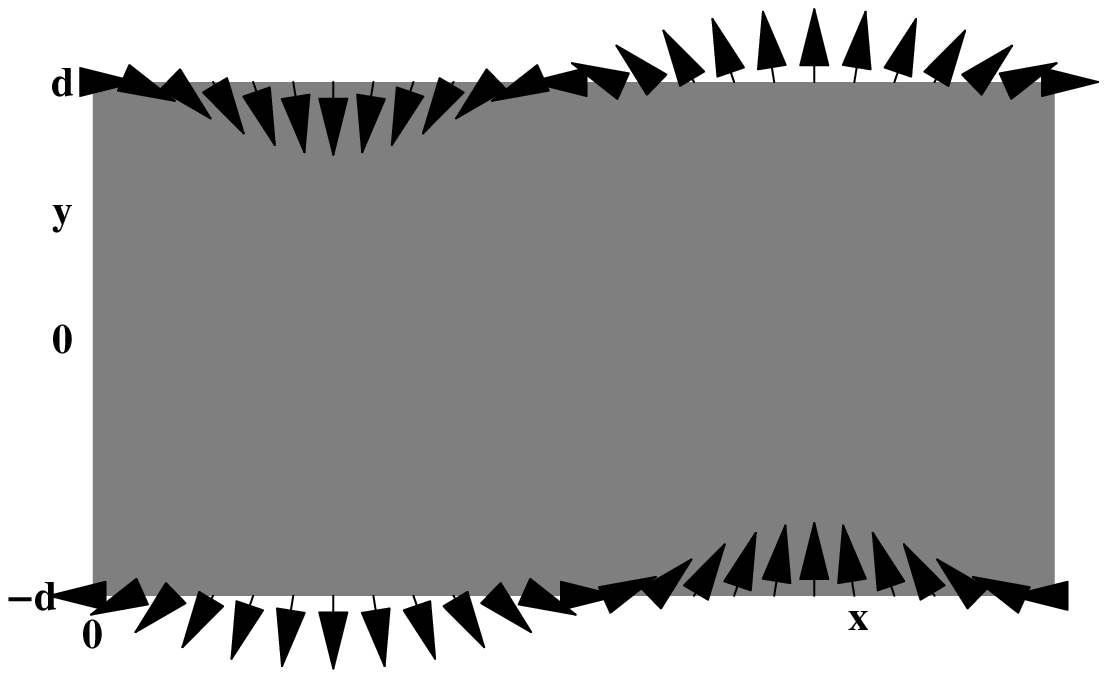}
   \put(-9.1,3.1){}
\put(-1.2,-.2){}
  \caption{}
\end{figure}
\newpage

\end{document}